\documentclass[final]{aipproc}
\layoutstyle{6x9}

\usepackage{amsfonts,amsmath,amssymb,amsthm,bbm}
\usepackage{graphicx
}

\newcommand{\C}{{\mathbb C}}

\newcommand{\R}{{\mathbb R}}

\newcommand{\cE}{{\mathcal E}}

\newcommand{\cH}{{\mathcal H}}

\newcommand{\SU}{\mathrm{SU}}

\newcommand{\U}{\mathrm{U}}

\newcommand{\id}{\mathbb{I}}

\def\cHNJ{\cH_N^{(J)}}
\def\cHN{\cH_N}

\newcommand{\be}{\begin{equation}}
\newcommand{\ee}{\end{equation}}
\newcommand{\beq}{\begin{eqnarray}}
\newcommand{\eeq}{\end{eqnarray}}
\newcommand{\bea}{\begin{eqnarray}}
\newcommand{\eea}{\end{eqnarray}}
\newcommand{\nn}{\nonumber}

\newcommand{\su}{{\mathfrak su}}

\renewcommand{\u}{{\mathfrak u}}

\newcommand{\la}{\langle}
\newcommand{\ra}{\rangle}

\newcommand{\f}{\frac}

\def\nn{\nonumber}
\def\pp{\partial}

\def\arr{\rightarrow}

\def\Ea{E^{(\alpha)}}
\def\Eb{E^{(\beta)}}

\def\dag{^\dagger}

\newcommand{\matr}[2]{\left(\begin{array}{#1}#2\end{array}\right)}

\begin{document}

\title{New tools for Loop Quantum Gravity with applications to a simple model}


\author{Enrique F. Borja}
{
address={Institute for Theoretical Physics III, University of Erlangen-N\"{u}rnberg, Staudtstra{\ss}e 7, D-91058 Erlangen (Germany).}
,altaddress={Departamento de F\'{\i}sica Te\'{o}rica and IFIC, Centro Mixto Universidad de Valencia-CSIC. Facultad de F\'{\i}sica, Universidad de Valencia, Burjassot-46100, Valencia (Spain).}
}

\author{Jacobo D\'{\i}az-Polo}{
address={Department of Physics and Astronomy, Louisiana State University.\! Baton Rouge, LA,\! 70803-4001.}
}

\author{ Laurent Freidel}{
address={Perimeter Institute for Theoretical Physics, 31 Caroline St N, Waterloo ON,\! Canada N2L\! 2Y5.}
}

\author{I\~{n}aki Garay}{
address={Departamento de F\'{\i}sica Te\'{o}rica, Universidad del Pa\'{\i}s
Vasco, Apdo. 644, 48080 Bilbao (Spain).}
,altaddress={Institute for Theoretical Physics III, University of Erlangen-N\"{u}rnberg, Staudtstra{\ss}e 7, D-91058 Erlangen (Germany).}
}

\author{Etera R. Livine}{
address={Laboratoire de Physique, ENS Lyon, CNRS-UMR 5672, 46 All\'ee d'Italie, Lyon 69007, France.}
,altaddress={Perimeter Institute for Theoretical Physics, 31 Caroline St N, Waterloo ON,\! Canada N2L\! 2Y5.}
}

\begin{abstract}
Loop Quantum Gravity is now a well established approach to quantum gravity. One of the main challenges still faced by the theory is constructing a consistent dynamics which would lead back to the standard dynamics of the gravitational field at large scales. Here we will review the recent $\U(N)$ framework for Loop Quantum Gravity and the new spinor representation (that provides a classical setting for the $\U(N)$ framework). Then, we will apply these techniques to a simple model in order to propose a dynamics for a symmetry reduced sector of the theory. Furthermore, we will explore certain analogies of this model with Loop Quantum Cosmology.
\end{abstract}

\maketitle


\begin{center}
\bf{Introduction}
\end{center}

Loop Quantum Gravity (LQG) proposes a canonical quantization for the kinematics of quantum gravity. The Hilbert space is generated by spin networks (states defined over oriented graphs whose edges are labeled by $\SU(2)$ irreducible representations and whose vertices are decorated with $\SU(2)$ invariant tensors, the intertwiners). Despite the several advances made, the implementation of the dynamics is still an open problem.

In this article we describe two new frameworks for LQG developed recently: the $\U(N)$ framework and the spinor representation \citep{un1,un2,un3,return,Freidel:2010bw,Livine:2011gp,Borja:2011pd}. We will apply these new frameworks to a simple model based on 2 nodes joined by an arbitrary number $N$ of links. Then, we will identify and explore a global symmetry that selects a homogeneous and isotropic sector of this system \citep{2vertex,BGV}. Finally, using the spinor representation, we will propose for this system an action with an interaction term which encodes the effective dynamics of the model.

\section{The $\U(N)$ framework}

The $\U(N)$ framework \citep{un1,un2, un3} is very well suited to  
study the Hilbert space of intertwiners with $N$ legs. The
basic tool of this framework is the Schwinger representation
of the $\su(2)$ Lie algebra in terms of a pair of harmonic
oscillators $a$ and $b$:
$$
J_z=\f12(a\dag a-b\dag b),\quad
J_+=a\dag b,\quad J_-=a b\dag\,.
$$
Labeling the $N$ legs with the index $i$, we identify $\SU(2)$ invariant operators acting on pairs of (possibly equal) legs $i,j$:
\be
E_{ij}=a\dag_ia_j+b\dag_ib_j, \quad (E_{ij}\dag=E_{ji}),\quad\qquad
F_{ij}=a_i b_j - a_j b_i,\quad (F_{ji}=-F_{ij}).\nn
\ee
The operators $E$ form a $\u(N)$-algebra and they also form a closed algebra together with the operators $F,F\dag$.
%
The spin $j_i$ (the $\SU(2)$ irrep. of the leg $i$) is identified geometrically as the area associated to the leg $i$ and the total energy $E=\sum_i E_i$ gives twice the total area $J=\sum_i j_i$ associated to the intertwiner. The $E_{ij}$-operators change the energy/area carried by each leg, but still conserving the total energy; while the operators
$F_{ij}$ (resp. $F\dag_{ij}$) will decrease (resp. increase) the
total area $E$ by 2.The operators $E_{ij}$ allow then to navigate from state to state within each subspace $\cHNJ$ of $N$-valent intertwiners with fixed
total area $J$; and the operators  $F\dag_{ij}$ and $F_{ij}$ allow
to go from one subspace $\cHNJ$ to the next $\cHN^{(J\pm 1)}$, thus
endowing the full space of $N$-valent intertwiners with a Fock space
structure with creation operators $F\dag_{ij}$ and annihilation
operators $F_{ij}$. 
Finally, it is worth pointing out that the operators $E_{ij},F_{ij},F\dag_{ij}$ satisfy certain quadratic constraints, which correspond to a matrix algebra \citep{2vertex}.

\section{The 2 vertex model and the global symmetry}

\begin{figure}
\includegraphics[height=35mm]{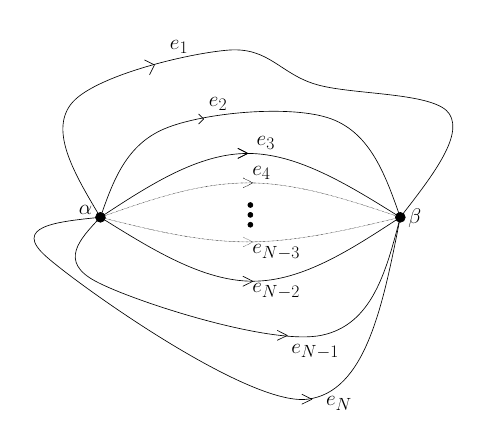}
\caption{The 2-vertex graph with vertices $\alpha$ and $\beta$ and the $N$ edges linking them.}
\label{2vertexfig}
\end{figure}

We consider the simplest class of non-trivial graphs for spin network states in LQG: a graph with two vertices linked by $N$ edges, as shown in fig.\ref{2vertexfig}.
There are matching conditions \citep{un2} imposing that each edge
carries a unique $\SU(2)$ representation (same spin seen from
$\alpha$ and from $\beta$). This translates into the equal energy condition 
$\cE_i\,\equiv\,\Ea_i -\Eb_i \,=\,0$.
These constraints $\cE_k$ turn out to be part of a larger $\U(N)$
symmetry algebra satisfied by the operators $\cE_{ij}\,\equiv\, \Ea_{ij}-\Eb_{ji}
\,=\,\Ea_{ij}-(\Eb_{ij})\dag$.
Now, one can show \citep{2vertex} that by imposing the global
$\U(N)$-invariance generated by the $\cE_{ij}$'s on our 2-vertex system, one obtains
a single state $|J\ra$ for each total boundary area $J$. Thus, the
$\U(N)$ invariance is restricting our system to states that are
homogeneous and isotropic (the quantum state is the same at every
point of space, i.e. at $\alpha$ and $\beta$, and all directions or
edges are equivalent).

\section{Spinors and effective dynamics}

Based on the Schwinger representation of $\SU(2)$ in terms of
harmonic oscillators, it is possible to give a representation of the
classical phase of LQG in terms of spinor variables
\citep{return,Livine:2011gp}. The quantization of this classical
system will lead us back to the $\U(N)$ framework for intertwiners.
Focusing on this classical system, we write an action principle with
an effective dynamics of the spinors for the symmetry reduced sector of the 2-vertex model.

Let us start by introducing the usual spinor notation. We define the spinor $z$ and its dual:
$$
|z\ra=\matr{c}{z^0\\z^1}\in\C^2, \qquad \la z|=\matr{cc}{\bar{z}^0
&\bar{z}^1}, \qquad |z]\equiv
\begin{pmatrix}-\bar{z}^1\\\bar{z}^0 \end{pmatrix}.
$$
In order to describe $N$-valent intertwiners, we consider $N$
spinors $z_i$ satisfying a closure condition\footnotemark\,that, in
terms of their components, is given by:
\begin{equation}
\sum_i |z_i\ra\la z_i|\propto\id\,\Leftrightarrow\,
\sum_i z^0_i\,\bar{z}^1_i=0,\quad \sum_i \left|z^0_i\right|^2=\sum_i
\left|z^1_i\right|^2=\f12\sum_i \la z_i|z_i\ra. \label{closure}
\end{equation}
Solutions are parameterized in terms of a positive number
$\lambda\in\R_+$ and a unitary matrix $u\in\U(N)$ up to
$\U(\!N\!-\!2\!)\!\times\!\SU(2)$ right-transformations with
$z_i^0=\!\sqrt{\lambda}\,u_{i1}$ and $z_i^1\!=\sqrt{\lambda}\,u_{i2}$.
\footnotetext{We associate to each spinor $z_i$ a 3-vector $\vec{X}(z_i)=\la z_i|\vec{\sigma}|z_i\ra$, with $\vec{\sigma}$ the Pauli matrices. Then the closure constraint is $\sum_i \vec{X}(z_i)=0$ and we identify $\vec{X}(z_i)$ as the normal vector to the dual surface to the leg $i$.}

The phase space is defined by the canonical Poisson bracket
$\{z^a_i,\bar{z}^b_j\}\,\equiv\,i\,\delta^{ab}\delta_{ij}\,$. The
quantization will be promoting $z_i$ and $\bar{z_i}$ as the
annihilation and creation operators of harmonic oscillators. Then
the classical matrices $M_{ij}=\la z_i |z_j \ra$ and $Q_{ij}=[z_j
|z_i\ra$ are the classical counterparts of the operators $E$ and
$F$.

The $\U(N)$-action  on spinors is the simple $N\times N$ matrix action $(Uz)_i=\sum_j U_{ij}z_j$. Defining the ``homogeneous cosmological'' sector as the $\U(N)$-invariant sector, satisfying $\la z_i^\alpha |z_j^\alpha \ra=\la z_i^\beta |z_j^\beta \ra$ and invariant under $z^\alpha,z^\beta\,\arr Uz^\alpha,\bar{U}z^\beta$, imposes that all the $\alpha$-spinors are equal to the $\beta$-spinors up to a global phase, $\bar{z}^{(\alpha)}_i\,=\,e^{i\phi}\,z^{(\beta)}_i$. And we get a reduced phase space with two parameters, the total area $\lambda$ and its conjugate angle $\phi$ encoding the curvature. Our ansatz for the dynamics of this ``cosmological'' sector is:
\be
S_{inv}[\lambda,\phi] \,=\, -2 \int dt\,\left(\lambda \pp_t \phi
-\lambda^2\left(\gamma^0-\gamma^+e^{2i\phi}
-\gamma^-e^{-2i\phi}\right)\right),
\ee
which corresponds to the quantum Hamiltonian obtained in \citep{2vertex} with the operators $E, F, F\dag$ ($\gamma^0, \gamma^\pm$ are coupling constants). In this classical case, the equations of motion can be solved exactly \citep{return} with certain interesting analogies with (the effective dynamics of) Loop Quantum Cosmology, showing that the dynamics of the $\U(N)$-invariant sector of the 2-vertex graph model can be interpreted as describing homogeneous and isotropic cosmology  (see \citep{Livine:2011up} for a recent discussion).

\section{Conclusions}

The $\U(N)$ framework and the spinor representation represent a new and refreshing way to tackle several important issues in Loop Quantum Gravity. In this work we have discussed these new frameworks and we have reviewed a proposal for the effective dynamics of the homogeneous and isotropic sector of the model at the classical level (that coincides with the quantum Hamiltonian proposed in \citep{2vertex}).  

We have described the main features of the $\U(N)$ framework, like the Fock space structure of the Hilbert space of intertwiners with $N$ legs and we further used this $\U(N)$ structure on the 2-vertex graph to define a symmetry reduction to the homogeneous and isotropic sector. We have then, using the spinor representation, introduced a Hamiltonian consistent with this symmetry reduction, which can be solved exactly and shown to have certain interesting analogies with the dynamics of Loop Quantum Cosmology.

\begin{center}
\bf{Acknowledgments}
\end{center}

This work was in part supported by the Spanish MICINN research
grants FIS2008-01980, FIS2009-11893 and ESP2007-66542-C04-01 and by
the grant NSF-PHY-0968871. IG is supported by the Department of
Education of the Basque Government under the ``Formaci\'{o}n de
Investigadores'' program.



\begin{thebibliography}{10}


\bibitem{un1}
F. Girelli and E.R. Livine. {\it Reconstructing Quantum Geometry
from Quantum Information: Spin Networks as Harmonic Oscillators},
Class.Quant.Grav. 22 (2005) 3295 [arXiv:gr-qc/0501075].

\bibitem{un2}
L. Freidel and E.R. Livine. {\it The Fine Structure of $\SU(2)$
Intertwiners from $\U(N)$ Representations}, J.Math.Phys. 51 (2010)
082502 [arXiv:0911.3553].

\bibitem{un3}
L. Freidel and E.R. Livine. {\it $\U(N)$ coherent states for Loop
Quantum Gravity}, J.Math.Phys. 52 (2011) 052502 [arXiv:1005.2090].

\bibitem{return}
E.F. Borja, L. Freidel, I. Garay and E.R. Livine. {\it U(N) tools
for Loop Quantum Gravity: The Return of the Spinor},
Class.Quant.Grav. 28 (2011) 055005 [arXiv:1010.5451].

\bibitem{Freidel:2010bw}
Laurent Freidel and Simone Speziale. {\it From twistors to twisted
geometries}, Phys.Rev. D82 (2010) 084041 [arXiv:1006.0199v1].

\bibitem{Livine:2011gp}
Etera~R. Livine and Johannes Tambornino. {\it Spinor Representation
for Loop Quantum Gravity}. [arXiv:1105.3385v1].

\bibitem{Borja:2011pd}
E.F. Borja, J. Diaz-Polo and I. Garay. {\it $\U(N)$ and holomorphic methods for LQG and Spin Foams}. [arXiv:1110.4578v1].

\bibitem{2vertex}
E.F. Borja, J. Diaz-Polo, I. Garay and E.R. Livine. {\it Dynamics
for a 2-vertex Quantum Gravity Model}, Class.Quant.Grav.27 (2010)
235010 [arXiv:1006.2451v2].

\bibitem{BGV}
E.F. Borja, I. Garay and F. Vidotto. {\it Learning about quantum gravity with a couple of nodes}. [arXiv:1110.3020v1].

\bibitem{Livine:2011up}
E.R. Livine and M. Martin-Benito. {\it Classical Setting and Effective Dynamics for Spinfoam Cosmology}. [arXiv:1111.2867v1].


\end{thebibliography}
\end{document}